\documentclass[amt, manuscript]{copernicus}

\begin{document}
\nolinenumbers
\title{Introduction to the calibration ringing effect in satellite hyperspectral atmospheric spectrometry}

\Author[1]{Dussarrat}{Pierre}
\Author[1]{Theodore}{Bertrand}
\Author[1]{Coppens}{Dorothee}
\Author[2]{Standfuss}{Carsten}
\Author[2]{Tournier}{Bernard}

\affil[1]{EUMETSAT, Eumetsat-Allee 1, 64295 Darmstadt, Germany}
\affil[2]{SPASCIA, 10 Avenue de l'Europe, 31520 Ramonville-Saint-Agne, France}

\correspondence{Pierre Dussarrat (Pierre.Dussarrat@eumetsat.int)}

\runningtitle{TEXT}

\runningauthor{TEXT}

\received{}
\pubdiscuss{} %% only important for two-stage journals
\revised{}
\accepted{}
\published{}

%% These dates will be inserted by Copernicus Publications during the typesetting process.

\firstpage{1}

\maketitle

\begin{abstract}
Atmospheric remote spectrometry from space has become in the last 20 years a key component of the Earth monitoring system: their large coverage and deci-kelvin stability have demonstrated their usefulness for weather prediction, atmospheric composition monitoring as well as climate monitoring. It is thus critical to investigate the possible sources of errors associated to this technique. One of them is the so-called "calibration ringing error" that appears in Fourier transform spectrometers at the radiometric calibration step when the instrument transmission varies at the scale of the spectral resolution and is not accounted by the data users. This paper exposes the theoretical basis of this particular type of radiometric uncertainty. Its sensitivity to instrumental parameters as well as the impact on the radiometrically calibrated measurements is assessed in the context of atmospheric infrared sounding using Fourier transform spectrometers. It is shown that this error is an intrinsic feature of such instruments that could safely be ignored in early-generation instruments but will have to be taken into account in the new generation ones as it can yield a significant degradation of the radiometric error budget.
\end{abstract}

%\copyrightstatement{TEXT}

\introduction  %% \introduction[modified heading if necessary]
Fourier transform spectrometry is a powerful mean to sound the atmosphere by resolving the spectral lines of the infrared radiation emitted by the Earth and the atmosphere.  The light enters the instrument which is mainly composed of a telescope and a two-arm interferometer. The interference pattern is recorded on the instrument detector in function of the difference of the two arms' lengths. The light spectrum is recovered by Fourier transform of the interferogram. Finally, the spectrum exhibits absorption and emission lines representative of the Earth atmosphere composition and thermodynamic state.

In order to gather as much information as possible, it is desirable that such instruments are covering the thermal infrared domain in large spectral bands. Unavoidably, measurements are acquired at wavelengths where the properties of the optical elements are not necessarily optimal: the optical transmission can exhibit local gradients or even oscillations. On the other hand, the desirable optimization of the spectral resolution of the instrument is capped by several technical reasons, as the interferometer mechanism capability, the space-to-Earth link finite bandwidth and the overall processing power. Finally, the instrument's spectral band and resolution choices are a result of a complex trade-off between the scientific interests and technical limitations. Thus, in general, we expect to perform measurements at wavelengths where the instrument's optical transmission varies within the domain of the spectral response function. 

Usually, the optical transmission on the band of interest is characterised on-board with specific calibration scheme using for example, on-board black-bodies and deep space measurements (see, {\it e.g.}, [\cite{Revercomb1988}] and [\cite{Tournier2002}]). Then, the optical transmission is removed from the raw Earth view measurements. However, we still expect the occurrence of high-frequency residual spectral modulations, often confused with a spectral noise. The effect of transmission induced biases was first reported by \cite{Revercomb2017} and designated as "true ringing". The prefix "true" was introduced to depict an effect inherent to the radiometric response function of the instrument as opposed to other identified processing errors. In the following, to be more rigorous and precise, we will use the "calibration ringing" denomination.

The desire to improve the spectrometers' radiometric accuracy below the deci-kelvin limit has recently brought attention to this particular effect in the satellite remote sensing community (see [\cite{Revercomb2017}] for the CrIS instrument and [\cite{Atkinson2018}]). In particular, the authors have been directly exposed to calibration ringing issues for the InfraRed Sounder (IRS) instrument on-board the future Meteosat Third Generation-Sounding (MTG-S) mission [\cite{Coppens2017}], for which both transmission gradients and modulations are expected. This specific analysis will be presented in a future paper. For now, we aim at presenting in a pedagogical way the theoretical origins of the calibration ringing to constrain its amplitude in the early development stages of the future missions and to pave the way to the development of correction schemes.

In the second section, the theoretical basis of the phenomenon is discussed and the instrumental parameters leading to calibration ringing error are outlined. Then, in section~3, simulations illustrating and confirming the theoretical results, in the particular case of thermal infrared Fourier spectrometers, are presented.

\section{Theory}
\subsection{Fourier transform spectrometers}
A Fourier transform spectrometer records interference patterns produced by a target source of spectrum $Sp$ in function of the two arms optical path length difference $x$, referred to in the following as OPD. The produced interferogram, omitting its baseline, is calculated as an integral over the positive wavenumbers (equal to one over the wavelength) of all spectral component interferences:
\begin{equation}
\label{eq:1}
I(x) = \int_{0}^\infty Sp(\nu)~cos[2\pi\nu x]~d\nu = \int_{-\infty}^\infty \hat{Sp}(\nu)~e^{2i\pi\nu x}~d\nu = FT^{-1}[\hat{Sp}(\nu)]
\end{equation}
where the symmetrical target spectrum is defined as $\hat{Sp}(\nu > 0) = Sp(\nu)/2$ and $\hat{Sp}(\nu < 0) = Sp(-\nu)/2$. Thus, the interferogram is directly related to the inverse Fourier transform of the symmetrical input spectrum and therefore the input spectrum can simply be retrieved by direct Fourier transform of the interferograms.

The interferogram is recorded on a finite OPD range and can also be smoothed by an apodisation function: $Apod(x)$. It includes self-apodisation effects, which we assume as either negligible or without spectral dependence, such that the apodisation function is the same for all wavenumbers of the retrieved spectrum. The direct Fourier transform of the apodisation function defines the spectral response function $SRF$ so that the retrieved spectrum $Sp_m$ at wavenumber of interest $\nu_0$ is equal to the input spectrum convoluted with the SRF:
\begin{equation}
\label{eq:2}
Sp_m(\nu_0) = FT[I(x) \times Apod(x)](\nu_0) = \left[\hat{Sp} \otimes SRF\right](\nu_0)
\end{equation}

The transmission of the optical system $R(\nu)$ is usually not perfectly constant and equal to one over the spectral range of interest, so the input spectrum should be multiplied by the transmission in the previous equations. The recorded interferogram becomes: $I_r(x) = FT^{-1}[\hat{Sp}.R]$. Assuming that a set of reference measurements such as on-board black body and deep space views are available to compute the optical transmission, we can perform a radiometric calibration (see [\cite{Revercomb1988}] and [\cite{Tournier2002}]) and the calibrated spectrum $Sp_r$ writes:
\begin{equation}
\label{eq:3}
Sp_r(\nu_0) = \frac{FT[I_r(x) \times Apod(x)](\nu_0)}{[R\otimes SRF](\nu_0)} = \frac{[(\hat{Sp}.R) \otimes SRF](\nu_0)}{[R\otimes SRF](\nu_0)}
\end{equation}

This equation forms the basis of a Fourier transform spectrometer calibration. Details of the processing have been omitted in order to focus on the specific calibration ringing error in the next sections. 

\subsection{Calibration ringing definition}
In the following, we show that the measurement SRF after radiometric calibration is distorted by the transmission oscillations and that users who omit these distortions in their applications will perceive it as a radiometric error, that we call a {\it calibration ringing error}.

Defining the centered and swapped transmission at the wavenumber $\nu_0$ of interest: $R_{\nu0}(\nu) = R(\nu_0- \nu)$, the calibrated spectrum for this wavenumber writes:
\begin{equation} \label{eq:5}
\frac{[(\hat{Sp}.R) \otimes SRF](\nu_0)}{[R\otimes SRF](\nu_0)} = \frac{\left[\hat{Sp} \otimes (R_{\nu0}(\nu) SRF(\nu))\right](\nu_0)}{[R\otimes SRF](\nu_0)} = \hat{Sp} \otimes \left[ \frac{R_{\nu0}(\nu)}{[R\otimes SRF](\nu_0)} \times SRF(\nu)\right](\nu_0)
\end{equation}

This notation brings forward the fact that the SRF is locally distorted due to the transmission fluctuations around $\nu_0$. The distorted SRF at $\nu_0$ is defined as:
\begin{equation} \label{eq:52}
\boxed{
SRF_{dis,\nu0}(\nu) = \frac{R_{\nu0}(\nu)}{[R\otimes SRF](\nu_0)} \times SRF(\nu)
}
\end{equation}

If the distortion is not corrected or accounted in the applications, the measurement differs from the expectation of the data users and we define the {\it calibration ringing error} as the difference between the measurement with distortions and the one without:
\begin{equation} \label{eq:5}
\boxed{
\Delta(\nu_0) = \left[\hat{Sp} \otimes  SRF_{dis,\nu0} - \hat{Sp} \otimes SRF \right](\nu_0) =
\hat{Sp} \otimes \left[ \left( \frac{R_{\nu0}(\nu)}{[R\otimes SRF](\nu_0)} - 1 \right) \times SRF(\nu)\right](\nu_0)
}
\end{equation}

From this equation, it is evident that the calibration ringing error disappears only if the transmission is flat at the scale of the SRF {\it i.e.} $R_{\nu0}(\nu) = [R\otimes SRF](\nu_0)$. If not, a residual error that is a function of the input spectrum and of the transmission around $\nu_0$ is expected. In other words, the instrument cannot disentangle the input spectrum information from the transmission fluctuations. 

We underline that this error can be avoided if the radiometric transmission and the related SRF distortions are explicitly considered by the data users, for instance in the radiative transfer models. However, in practice, it appears computationally heavy and this means that each detector of a spectro-imager has to be processed as a self-standing instrument, which is considered as unfeasible by most of the users of the community, all the more in view of near real time processing. 

Finally, we stress that the calibration ringing error should not be confused with the nominal appearance of oscillating side lobes in the SRF of weakly apodized Fourier transform spectrometers or with other artificial ringing effects caused by any Fourier transform operations within the processing chain through improper smoothing of the band edges (assigned to Gibbs effects).

\subsection{Framework}
In order to gain insight on the calibration ringing phenomenon and assess its impact on the calibrated spectra, we have set up a theoretical framework in which we have assumed that the interferogram edges are smoothed by applying an apodisation function defined by the convolution of a Gaussian of standard deviation $\sigma_x$ with a door whose width is slightly smaller than the maximum OPD, noted $OPD_m$, by a margin $\delta_x$, with $\sigma_x = \delta_x/2$:
\begin{equation} \label{eq:61}
\begin{split}
Apod(x) &= [door[x,2(OPD_m -\delta x)] \otimes gauss[x,\sigma_x]]\\
SRF(\nu) &= FT\left[Apod(x)\right] = 2(OPD_m - \delta x) \times sinc[2\pi\nu (OPD_m - \delta x)] \times exp[-\sigma^2_x\nu^2/2]
\end{split}
\end{equation}
In practice, no interferogram sample is recorded above the maximum OPD, therefore the apodisation should be perfectly zero for $|x| > OPD_m$, which disagrees with this framework (see Eq.~\ref{eq:61}). Since the author's objective is to grant physical insight of the phenomenon, we keep this simplification to preserve analytical results. Moreover, using $\sigma_x = \delta_x/2$ ensures that the amplitude of the apodisation function is very small above $OPD_m$, which supports the approximation.

It is worth recalling that the spectral Nyquist sampling step, that is to say the minimum spectral sampling step achievable, is defined as $\delta\nu = 1/(2OPD_m)$ and the sampling frequency is then: $1/\delta\nu = 2OPD_m$.

\subsection{Gradient}\label{section:theory_gradient}
We consider in this section that the transmission $R(\nu)$ varies linearly around $\nu_0$. Using the first local derivative $R'(\nu_0)$, the transmission writes: $R(\nu_0 + \nu) = R(\nu_0) + R'(\nu_0) \times \nu$ and thus :$R_{\nu0}(\nu) = R(\nu_0) - R'(\nu_0) \times \nu$. In that case the calibration term $[R\otimes SRF](\nu_0)$ is simply equal to $R(\nu_0)$. 

Plugging eq. \ref{eq:61} into eq. \ref{eq:52} and using the gradient model, we obtain:
\begin{equation} \label{eq:7}
\boxed{
SRF_{dis,\nu0}(\nu) = SRF(\nu) - \frac{R'(\nu_0)}{\pi R(\nu_0)} \times sin[2\pi\nu (OPD_m - \delta x)] \times exp[-\sigma^2_x\nu^2/2]
}
\end{equation}
Thus, in the presence of a gradient, the distorted SRF exhibits an oscillating term. The spectral extension of the distortion term around $\nu_0$ is equal to $1/\sigma_x$. Therefore, in case of a very weak apodisation, $\sigma_x \rightarrow 0$, the distortion impacts the SRF over a large spectral range meaning that the calibration ringing error propagates over channels far away from $\nu_0$. On the contrary, a strong apodisation will limit the extension of the distortion term and will preserve the locality of the effect on the spectrum. 

The calibration ringing error then writes:
\begin{equation} \label{eq:71}
\boxed{
\Delta_g(\nu_0) =  - \frac{R'(\nu_0)}{\pi R(\nu_0)} \times \left[\hat{Sp} \otimes \left(sin[2\pi\nu (OPD_m - \delta x)] \times  exp[-\sigma^2_x\nu^2/2] \right)\right](\nu_0)
}
\end{equation}
The calibration ringing amplitude is thus proportional to the relative transmission derivative around $\nu_0$. The error shape is driven by the convolution of the input spectrum with a sine function at the frequency: $OPD_m - \delta x$ multiplied by a Gaussian function of width $1/\sigma_x$. The error is maximum around $\nu_0$ if the input spectrum locally oscillates approximately as a sine function at the $OPD_m - \delta x$ frequency.

Moreover, the ringing follows approximately the relation : $\Delta_g(\nu_0 + n/(OPD_m - \delta x)) = \Delta_g(\nu_0)$ for $n \in \mathbb{Z}$ and $ |n/(OPD_m - \delta x)| << 1/\sigma_x$. Therefore, the error is locally modulated with a $OPD_m - \delta x$ frequency, which is below or equal to half the sampling frequency. The error shape is characterized by alternating signs for adjacent samples and amplitudes depending on the spectrum shape in a spectral domain corresponding to a correlation length approximately equal to $1/2\sigma_x$. In the extreme case of unapodised spectra, the modulation frequency is equal to half of the sampling frequency, the convolution term in Eq.~9 becomes entirely independent of the local spectrum features around $\nu_0$, and the error shape is described by a zig-zag pattern of constant amplitude over large parts of the spectral band (in practice limited by the assumption of a constant relative transmission derivative, not applicable to the entire spectral band). The oscillating behavior of the error is at the origin of the "ringing" denomination.

\subsection{Modulation}\label{section:theory_modulation}
As a second step, we consider the case of an optical transmission exhibiting a local modulation of amplitude $\gamma$ at frequency $f$: $R(\nu) = R_0 + \gamma~cos[2\pi \nu f]$, thus $R_{\nu0}(\nu) = R_0 + \gamma~cos[2\pi(\nu - \nu_0) f]$. Such a modulation can be produced by an "etalon" effect {\it i.e.} multiple spurious reflections on the instrument optical interfaces (including the detector structure). Since the optical interfaces of an instrument cannot be perfectly coated, such an effect is expected to occur in most spectrometers. 

The calibration term $[R\otimes SRF](\nu_0)$ is in that case equal to $R_0 + \gamma~cos[2\pi\nu_0 f] \times Apod(f)$. It is then possible to compute the distorted SRF plugging Eq.~\ref{eq:61} into Eq.~\ref{eq:52} and using the modulation model. We also choose to present its approximation for $\gamma/R_0 << 1$ to extract the distortion term:
\begin{equation} \label{eq:72}
\boxed{
\begin{split}
SRF_{dis,\nu0}(\nu) &= \frac{R_0 + \gamma~cos[2\pi(\nu - \nu_0)f]}{R_0 + \gamma~cos[2\pi\nu_0f]~Apod(f)} \times SRF(\nu)\\
&\cong SRF(\nu) + \frac{\gamma}{R_0} \left( cos[2\pi(\nu - \nu_0)f] - cos[2\pi\nu_0 f] ~ Apod(f) \right) \times SRF(\nu)
\end{split}
}
\end{equation}
The distorted SRF exhibits an additional terms oscillating at the etalon frequency $f$. The distortion is nevertheless proportional to the ideal SRF; this preserves the locality of the distortion as opposed to the result obtained in the previous section (Eq.~8). 

Equation~\ref{eq:72} also shows that the distortion becomes negligible when the etalon frequency $f$ is very small compared to the sampling frequency $2OPD_m$. However, no damping of the distortion amplitude is expected for etalon frequencies above $OPD_m$. One can also conceive the presence of constructive interferences between the transmission terms and the SRF side lobes in Eq.~\ref{eq:72} for etalon frequency close to $OPD_m$, this particular resonance is confirmed by simulation in section~3.3.

The calibration ringing error then becomes:
\begin{equation} \label{eq:73}
\boxed{
\Delta_m(\nu_0) \cong  \frac{\gamma}{R_0} \times\left[\hat{Sp} \otimes  \left[\left( cos[2\pi(\nu_0 - \nu)f] - cos[2\pi\nu_0 f]\times Apod(f)  \right)  \times SRF(\nu)\right]\right](\nu_0)
}
\end{equation}
The amplitude of the calibration ringing error is proportional to the relative amplitude of the transmission modulation $\gamma/R_0$. The complex beating of the SRF and the etalon components in Eq.~\ref{eq:73} and its convolution with the input spectrum will lead to an impression of chaotic noise. Finally, in opposition to the previous section, the spectral extension of the error is given here by the SRF spectral range itself and therefore will stay local.

We underline that in the two previous cases if the transmission drops close to zero the calibration ringing error becomes heavily sensitive to any RTF variations, it is very relevant in the case of transmission cut-offs that could occur at the edges of the spectral bands.

\section{Simulation}
In order to assess the pertinence of the previously obtained theoretical characteristics of the calibration ringing error, we have simulated the occurrence of the error in different ideal cases exhibiting transmission gradients and modulations.

\begin{figure*}
\includegraphics[width=16cm]{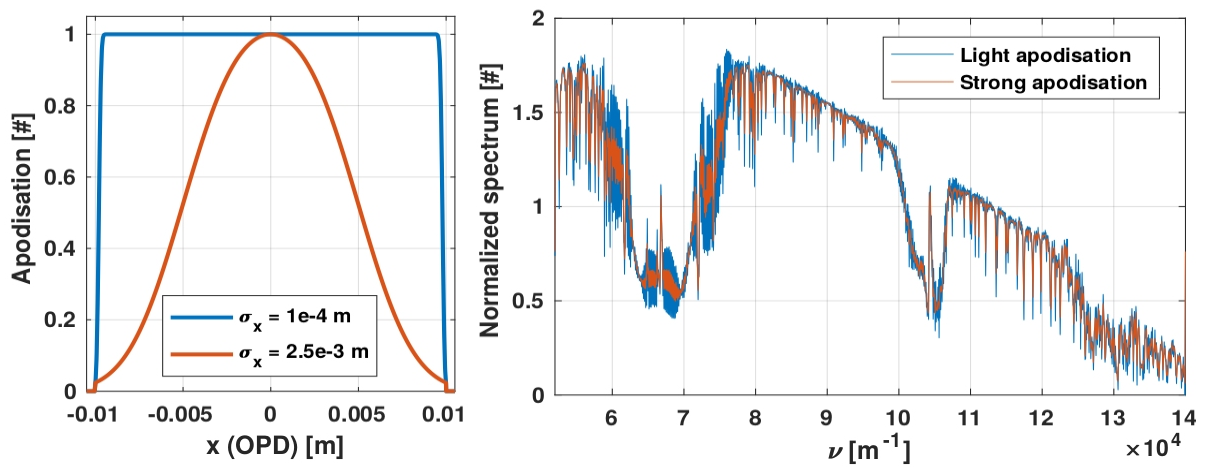}
\caption{The light (blue) and strong apodisation (red) are represented in function of the OPD on the left panel, the smoothing parameter $\sigma_x$ is equal to $10^{-4}~ m$ and $2.5~10^{-3}~m$ respectively. The input spectrum measured with a 1~cm Fourier transform interferometer and normalized is computed for both apodisations and represented in function of the wavenumber on the right panel.}
\label{fig:Apod}
\end{figure*}

We have arbitrarily chosen a high resolution (0.1~m$^{-1}$) spectrum generated by collaborators at ECMWF (European Centre for Medium-Range Weather Forecasts) using a line-by-line radiative transfer model over a realistic scene including the effect of all main trace gases in the thermal infrared (Fig.~\ref{fig:Apod}). For simplicity, the amplitude of this input spectrum is normalized to have a mean value equal to one; this implies that the calibration ringing error on the graphs can be directly interpreted as relative error.

We have considered an ideal spectrometer with a maximum OPD of 1~cm. We apply an apodisation function as defined in eq.~\ref{eq:61} with either $\sigma_x = \delta_x/2 = 10^{-4}$~m or $\sigma_x = 2.5~ 10^{-3}$~m (referred to as "light" or "strong" apodisation). In addition, the apodisation function is taken exactly null outside $\pm 1$~cm, since in practice, no samples are recorded over the maximum OPD. We recall that the theoretical framework used in the previous sections is not perfectly canceling the apodisation above maximum OPD (see section 2.3).

The interferogram that would be recorded by the instrument, $\hat{Sp}.R \otimes SRF$, has been computed as a discrete sum of all input spectral components and the measured spectrum is recovered by inverse Fourier transform as done in the processing of actual measurements from Fourier transform spectrometers. The same method has been applied to generate the {\it undistorted} spectrum $\hat{Sp} \otimes SRF$ and the radiometric calibration factor $R \otimes SRF$:
\begin{equation} \label{eq:79}
\begin{split}
&\hat{Sp}.R \otimes SRF = FT^{-1}\left[ \left(\sum_\nu Sp(\nu) \times R(\nu) \times cos[2\pi\nu x] \right)(x)\times Apod(x) \right]\\
&\hat{Sp} \otimes SRF = FT^{-1}\left[ \left(\sum_\nu Sp(\nu) \times cos[2\pi\nu x] \right) (x) \times Apod(x) \right]\\
&R \otimes SRF = FT^{-1}\left[ \left(\sum_\nu R(\nu) \times cos[2\pi\nu x]\right) (x) \times Apod(x) \right]
\end{split}
\end{equation}

We have over-sampled the retrieved spectrum by applying a zero-padding (until $x =\pm 5$~m) at interferogram level in order to generate highly resolved plots (at 0.1~m$^{-1}$ resolution) which is useful for the comparison with the theory. In practice, in most processing, the spectra are computed at Nyquist sampling (equal to 50~m$^{-1}$ here) to optimize the data size. The oversampling has no impact on the performance assessment.

In the following, the computations are performed on the spectral range $5.5$ to $14~10^4$~m$^{-1}$ and the results represented between $7$ and $12~10^4$~m$^{-1}$. Moreover, the transmission edges are smoothed to avoid Gibbs effects at the computation window limits.

\subsection{Gradients}
\begin{figure*}
\includegraphics[width=16cm]{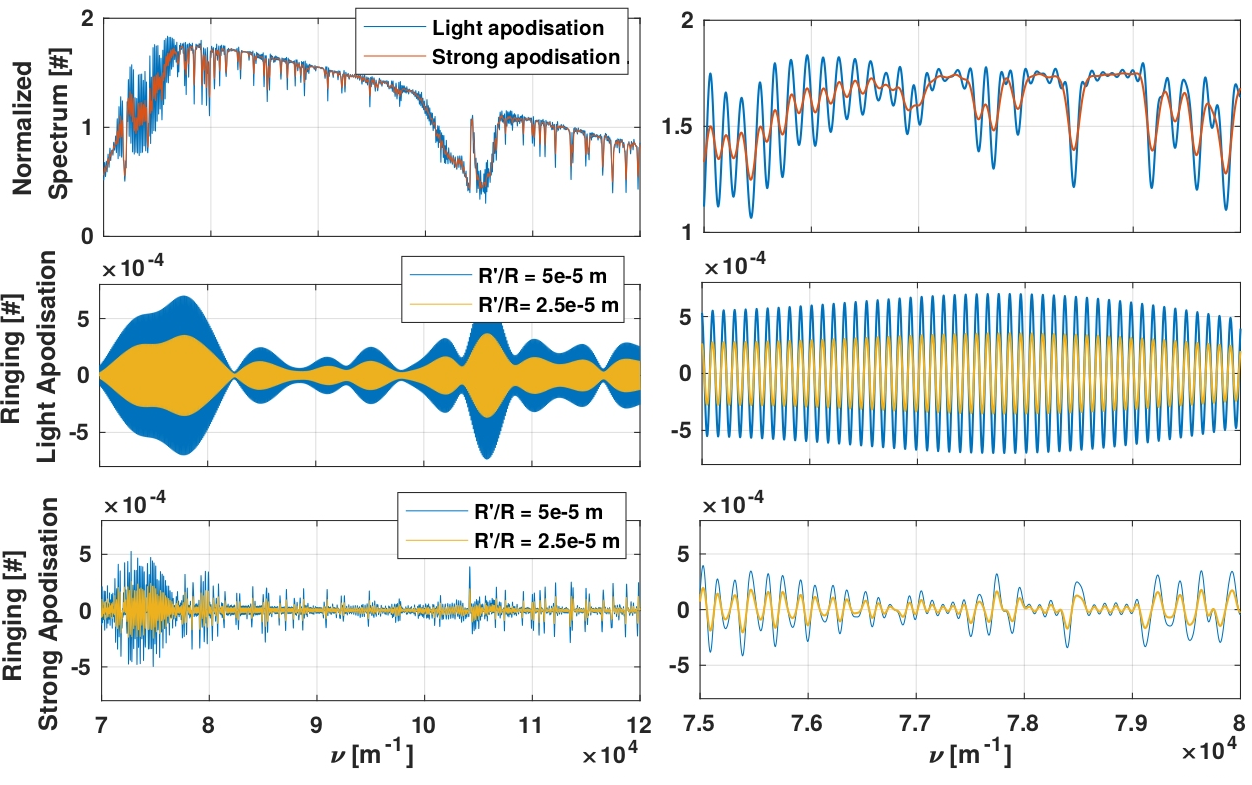}
\caption{The spectrum is computed with both apodisations and represented in the upper panels. The calibration ringing error is computed for the two apodisations and two relative transmission ratio $R'(\nu)/R(\nu) = 2.5~10^{-5}$~m and $5~10^{-5}$~m (yellow and blue curves respectively) in middle and bottom panels. The curves are presented for the whole spectral range (left) and zoomed in the range $7.5$-$8~10^4$~m$^{-1}$ (right).}
\label{fig:Gradient}
\end{figure*}

Figure~\ref{fig:Gradient} shows the simulated calibration ringing error induced by transmission relative gradients equal to $2.5~10^{-5}$~m and $5~10^{-5}$~m. To preserve a constant relative gradient on the whole band, the transmissions are taken as exponential curves.

The calibration ringing error appears as a modulation at frequency close to half the Nyquist frequency and proportional to the transmission slopes as expected from Eq.~\ref{eq:71}. Moreover, we also verify that the correlation length of the error is approximately proportional to $1/2\sigma_x$:
 \begin{itemize}
  \item Light apodisation: the correlation length is $5~10^3$~m$^{-1}$ which corresponds to 100 samples for Nyquist sampling,
  \item Strong apodisation: the correlation length is $2~10^2$~m$^{-1}$ which corresponds to 4 samples for Nyquist sampling.
\end{itemize}
These characteristic signatures are useful to recognize the calibration ringing error produced by gradients in the current generation of instruments from other types of biases.

The amplitude of the calibration ringing is larger in the case of a light apodisation since the error associated to each absorption line present in the spectrum is not local, as discussed in section~\ref{section:theory_gradient}, therefore the error is propagated to the adjacent lines. That effect generates constructive interferences and amplifies the maximal errors.

\subsection{Modulations}

In the second simulation we have considered a modulated optical transmission with a etalon frequency equals to 0.5~cm and two amplitudes, 1 and 2~$\%$. 

\begin{figure*}
\includegraphics[width=16cm]{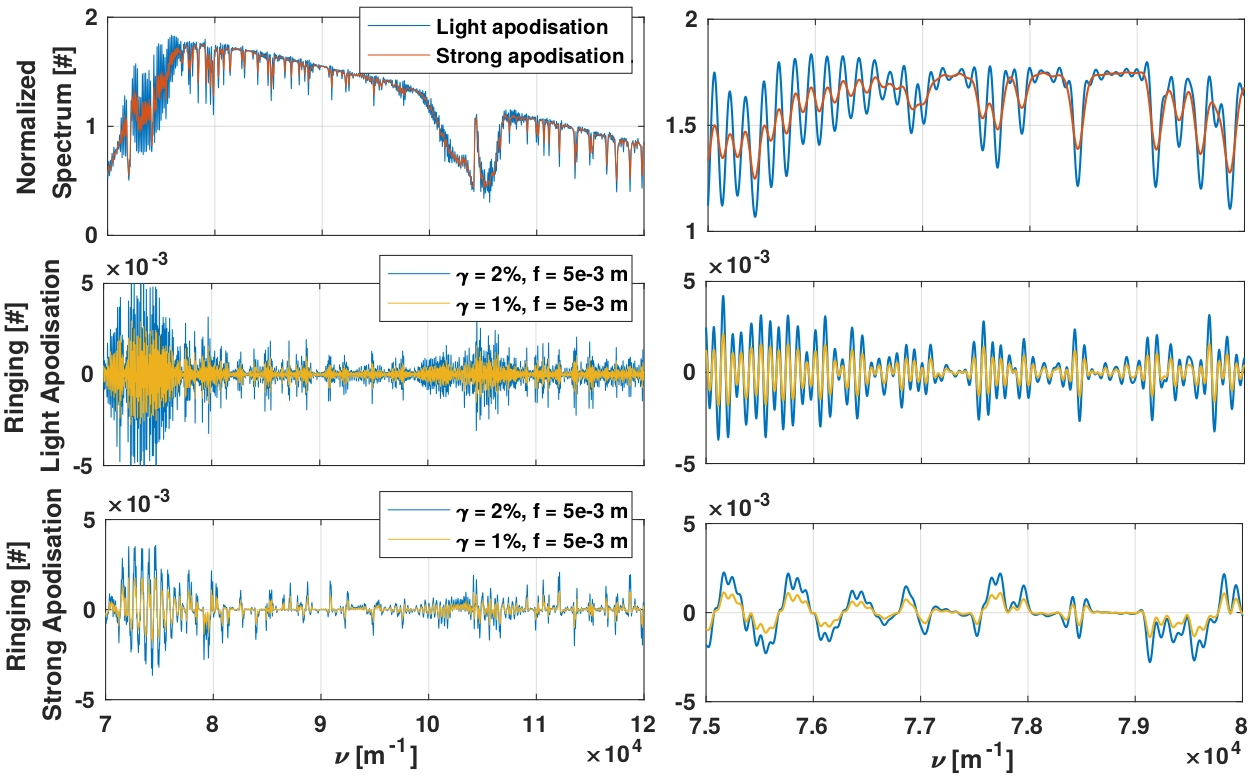}
\caption{The spectrum is computed with both apodisations and represented in the upper panels. The calibration ringing error is computed for the two apodisations, a modulation frequency equals to $5~10^{-3}$m and two amplitudes $\gamma = 1\%$ and $2\%$ (yellow and blue curves respectively) in middle and bottom panels. The curves are presented for the whole spectral range (left) and zoomed in the range $7.5$-$8~10^4$~m$^{-1}$ (right).}
\label{fig:Mod}
\end{figure*}

The resulting calibration ringing errors are shown on Fig.~\ref{fig:Mod} for both the light and strong apodisation cases. As expected from Eq.~\ref{eq:73}, the error is proportional to the etalon amplitude. Furthermore, as discussed in section~\ref{section:theory_modulation}, the calibration ringing is indeed localized around the absorption lines whatever the apodisation function used.

The singular behaviour of the error around the absorption lines is related to the complex interactions of the modulation terms in Eq.~\ref{eq:73}. Thus, we witness modulations at high frequency and an overall chaotic behaviour. 

It is worth noting that a moderate amplitude of the transmission modulation of only 2~$\%$ produces a relative error that already exceeds 0.005 for a light apodisation (equivalent signal-to-noise ratio of 200), which is significant compared to nowadays radiometric accuracy requirements (close to an equivalent signal-to-noise ratio of 100).

\subsection{Calibration ringing maximum}
Finally, to evaluate the impact of ringing on the radiometric calibration, we have computed the maximum error in equivalent black body temperature at 280~K between $7$ and $12~10^4$~m$^{-1}$ for many relative gradients and modulation frequencies with 1~$\%$ amplitude.

\begin{figure*}
\includegraphics[width=16cm]{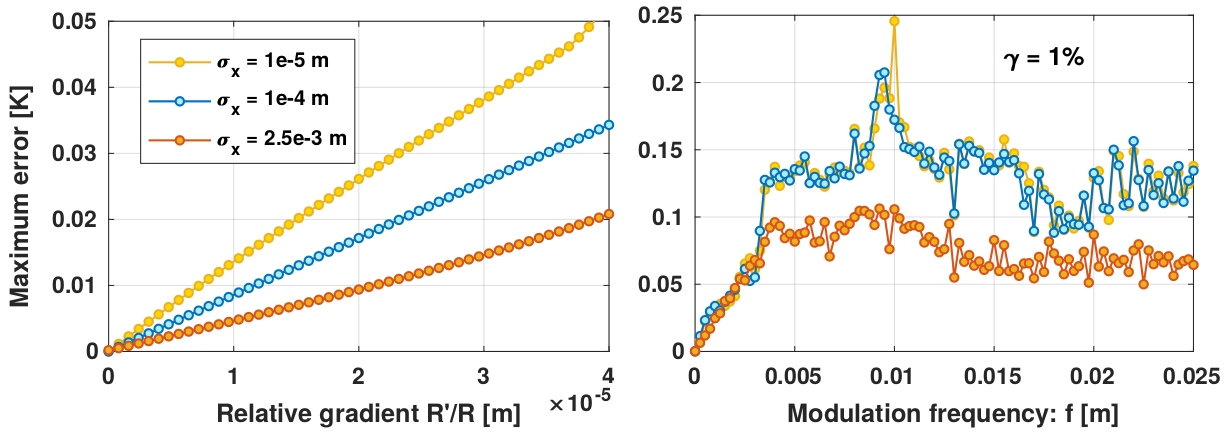}
\caption{Simulation of the maximum calibration ringing noise in equivalent temperature at 280 K in the band $7$ to $12~10^4$~m$^{-1}$ as a function of the transmission relative gradient amplitudes (left) and modulation frequencies with a $1\%$ amplitude (right). The errors are computed for three apodisation strengths: no, light and strong apodisation, with respectively $\sigma_x = 10^{-5}$, $10^{-4}$ and $2.5~10^{-3}$m.
}
\label{fig:OPD}
\end{figure*}

 On top of the light and strong apodisation cases, we have added the case "no apodisation" defined with $\sigma_x = 10^{-5}$~m. The calibration ringing error results are converted in equivalent temperature error dividing by the derivative of the Planck black body relation $\Gamma(\nu, T_{ref})$, at a reference temperature $T_{ref}$ of 280K. The maximum temperature errors on the band are computed and represented on Fig.~\ref{fig:OPD}.

The left panel of Fig.~\ref{fig:OPD} shows that the maximum calibration ringing is proportional to the relative gradient, as expected from Eq. 9. Moreover it is highly dependent on the apodisation strength ; in the cases "light" and "no apodisation", the error becomes highly non-localised and can interfere constructively, thus increasing the maximal error as discussed in section~\ref{section:theory_gradient}.

In the case of a modulation of the transmission (right plot of Fig.~\ref{fig:OPD}), the maximum amplitude of the error depends linearly on the etalon frequency for low values of $f$ and does not show a dependency on the apodisation. This a consequence of the fact that the spectral extension of the error is completely driven by the spectral width of the SRF as noted in section~\ref{section:theory_modulation}. At high frequencies, on the other hand, the maximum error reaches a plateau whose level depends on the strength of the apodisation. Between these two regimes, around the maximum OPD (equal to 1cm here), the error exhibits a large sensitivity to the etalon frequency, an effect that is damped in the presence of a strong apodisation. Moreover, we also verify that the error is proportional to the modulation amplitude.

In conclusion, Fig.~\ref{fig:OPD} sums up all keys parameters that can lead to calibration ringing error and particularly shows an overall positive impact of apodisation on the maximum error. These results pertain to the currently flying CrIS instrument and the future MTG-IRS, whose maximum OPD are 8 and 8.25~mm, respectively.

%\section{Application to current and future instruments ?}

\conclusions  %% \conclusions[modified heading if necessary]
We have presented in this paper a theoretical basis for the calibration ringing error that appears in Fourier transform spectrometers. We have shown that such an error is an intrinsic feature of instruments in which the transmission varies at the scale of the spectral resolution. It is worth noting at this point that the radiometric transfer functions or the set of distorted SRFs could be distributed along with the measurements for all spectral channels in order to be accounted for in the level-2 processing. This is however technically out of reach for most of the data users as this possibly represents a huge amount of data and is too processing intensive for now, all the more in view of near real time processing (and the potential need to update relevant instrument characteristics in case of temporal instability).

The calibration ringing error could be ignored in earlier interferometric IR atmospheric sounding missions for various reasons, spanning from a lower level of radiometric accuracy requirement to the processing strategy. In the case of IASI for instance, the RTF is rather flat, limiting the error by itself. Moreover, several processing characteristics (such as the numerical apodisation and the band merging) as well as a relatively strong self-apodisation render any residual bias undetectable as such. But with the emergence of new generation instruments, this effect has started attracting attention. In particular, the calibration ringing error has been identified as potentially relevant in the early development phases of the Infra-Red Sounder (IRS) onboard the Meteosat Third Generation (MTG) satellites.

Thus, a processing aiming at removing the ringing error is currently under evaluation by the authors and will be presented in a future paper. Moreover, in the last few years, \cite{Revercomb2017} and \cite{Atkinson2018} have also proposed correction prototypes and \cite{crisattheirstopicalmeeting} are currently developing a correction algorithm focusing on the CrIS instrument. The community has also begun the discussions to incorporate the instrument radiometric transfer function directly into the radiative transfer models to avoid such corrections.

In conclusion, it appears that the research is very active on the topic of calibration ringing in the community. The authors hope that the pedagogical theoretical work and the simulations presented in this paper properly describe the calibration ringing origins and will guide the development of the next innovations in the field.

\bibliographystyle{copernicus}
\bibliography{hsirl1.bib}

\end{document}